\documentclass[a4paper,twocolumn,superscriptaddress,11pt,accepted=2018-01-21]{quantumarticle}
\usepackage[utf8]{inputenc}
\usepackage[english]{babel}
\usepackage[T1]{fontenc}
\usepackage[numbers]{natbib}
\usepackage{float}

\usepackage{graphicx}% Include figure files
\usepackage{dcolumn}% Align table columns on decimal point
\usepackage{bm}% bold math
\usepackage{bbm}
\usepackage{braket}
\usepackage{amsmath}
\usepackage{amssymb}
\usepackage{enumerate}
\usepackage{booktabs}
\usepackage{listings}
\usepackage{subcaption}
\usepackage[justification=raggedright]{caption}
\usepackage{hyperref}% add hypertext capabilities
\usepackage{xcolor}

\makeatletter
\def\hlinewd#1{%
\noalign{\ifnum0=`}\fi\hrule \@height #1 %
\futurelet\reserved@a\@xhline}
\makeatother

\definecolor{dblue}{rgb}{0,0,0.7}
\definecolor{lgray}{rgb}{.4,.4,.4}
\begin{document}
\graphicspath{{./images/}{./images/plots/}}

\title{ProjectQ: An Open Source Software Framework for Quantum Computing}% Force line breaks with \\
%\thanks{A footnote to the article title}%

\author{Damian S. Steiger}
\orcid{0000-0003-1588-8930}
%\email{dsteiger@phys.ethz.ch}
\affiliation{Institute for Theoretical Physics, ETH Zurich, 8093 Zurich, Switzerland}

\author{Thomas H\"aner}
\orcid{0000-0002-4297-7878}
%\email{haenert@phys.ethz.ch}
\affiliation{Institute for Theoretical Physics, ETH Zurich, 8093 Zurich, Switzerland}

\author{Matthias Troyer}
\orcid{0000-0002-1469-9444}
%\email{troyer@phys.ethz.ch}
\affiliation{Institute for Theoretical Physics, ETH Zurich, 8093 Zurich, Switzerland}

\date{\today}% It is always \today, today,
             %  but any date may be explicitly specified

\begin{abstract}
We introduce ProjectQ, an open source software effort for quantum computing. The first release features a compiler framework capable of targeting various types of hardware, a high-performance simulator with emulation capabilities, and compiler plug-ins for circuit drawing and resource estimation. We introduce our Python-embedded domain-specific language, present the features, and provide example implementations for quantum algorithms. The framework allows testing of quantum algorithms through simulation and enables running them on actual quantum hardware using a back-end connecting to the IBM Quantum Experience cloud service. Through extension mechanisms, users can provide back-ends to further quantum hardware, and scientists working on quantum compilation can provide plug-ins for additional compilation, optimization, gate synthesis, and layout strategies.
\end{abstract}

\maketitle

%\tableofcontents

\section{Introduction}

Quantum computers are a promising candidate for a technology which is capable of reaching beyond exascale performance. There has been tremendous progress in recent years and we soon expect to see quantum computing test beds with tens and hopefully soon hundreds or even thousands of qubits. As these test devices get larger, a full software stack for quantum computing is required in order to accelerate the development of quantum software and hardware, and to lift the programming of quantum computers from specifying individual quantum gates to describing quantum algorithms at higher levels of abstraction.

The open source effort ProjectQ aims to improve the development of practical quantum computing in three key areas. First, the development of new quantum algorithms is accelerated by allowing to implement them in a high-level language prior to testing them on efficient high-performance simulators and emulators. Second, the modular and extensible design encourages the development of improved compilation, optimization, gate synthesis and layout modules by quantum computer scientists, since these individual components can easily be integrated into ProjectQ's full stack framework, which provides tools for testing, debugging, and running quantum algorithms. Finally, the back-ends to actual quantum hardware -- either open cloud services like the IBM Quantum Experience~\cite{ibmqewebsite} or proprietary hardware -- allow the execution of quantum algorithms on changing quantum computer test beds and prototypes. Compiling high-level quantum algorithms to quantum hardware will facilitate hardware-software co-design by giving feedback on the performance of algorithms: theorists can adapt their algorithms to perform better on quantum hardware and experimentalists can tune their next generation devices to better support common quantum algorithmic primitives.

\begin{figure}[t]
	\includegraphics[width=\linewidth]{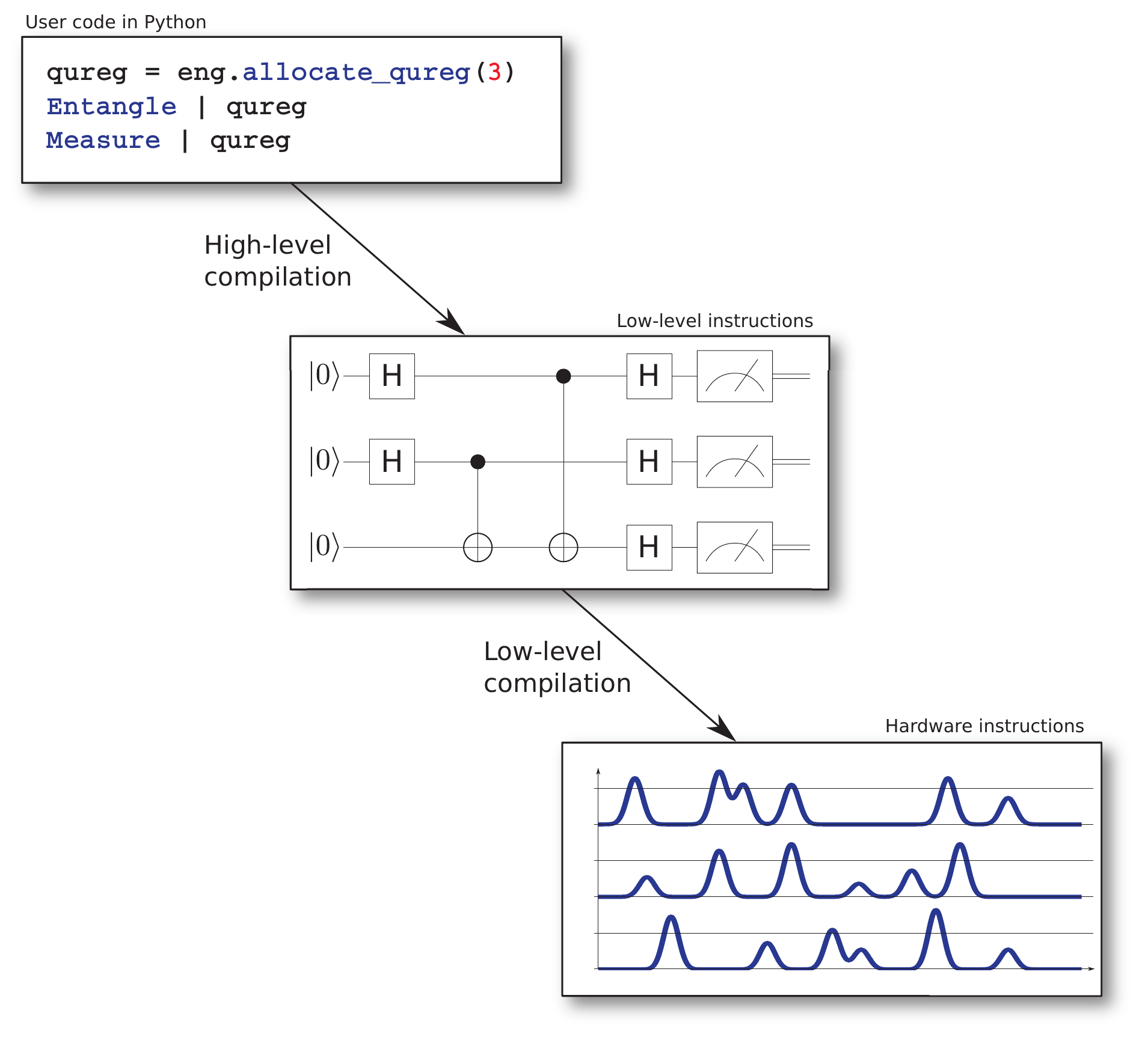}
	\caption{High-level picture of what a compiler does: It transforms the high-level user code to gate sequences satisfying the constraints dictated by the target hardware (supported gate set, connectivity, ...) while optimizing the circuit. The resulting low-level instructions are then translated to, e.g., pulse sequences.}
	\label{fig:hlcompiler}
\end{figure}

We propose to use a device independent high-level language with an intuitive syntax and a modular compiler design, as discussed in Ref.~\cite{haener2016}. The quantum compiler then transforms the high-level language to hardware instructions, optimizing over all the different intermediate representations of the quantum program, as depicted in Fig.~\ref{fig:hlcompiler}.
Programming quantum algorithms at a higher level of abstraction results in faster development thereof, while automatic compilation to low-level instruction sets allows users to compile their algorithms to any available back-end by merely changing one line of code. This includes not only the different hardware platforms, but also simulators, emulators, and resource estimators, which can be used for testing, debugging, and benchmarking algorithms. Moreover, our modular compiler approach allows fast adaptation to new hardware specifications in order to support all qubit technologies currently being developed.

Our high-level quantum language is implemented as a  domain-specific language embedded in Python; see the following code for an example:
\begin{lstlisting}[label={lst:add},numbers=none]
def AddConstant(eng, quint, c):
	with Compute(eng):
		QFT | quint
	
	# addition in the phases:
	phi_add(quint, c)
	
	Uncompute(eng)
\end{lstlisting}

To enable fast prototyping and future extensions, the compiler is also implemented in Python and makes use of the novel meta functions introduced in Ref.~\cite{haener2016} in order to produce more efficient code. Having the entire compiler implemented in Python is sufficient and preferred for current and near-term quantum test beds, as Python is widely used and allows for fast prototyping. If certain compiler components prove to be bottlenecks, they can be moved to a compiled language such as C++ using, e.g., \texttt{pybind} \cite{pybind}. Thus, ProjectQ is able to support both near-term testbeds and future large-scale quantum computers.

As a back-end, ProjectQ integrates a quantum emulator, as first introduced in Ref.~\cite{haenerSC2016}, allowing to simulate quantum algorithms by taking classical shortcuts and hence obtaining speedups of several orders of magnitude. For the simulation at a low level, we include a new simulator which outperforms all other available simulators, including its predecessor in Ref.~\cite{haenerSC2016}. Furthermore, our compiler has been tested with actual hardware and one of our back-ends allows to run quantum algorithms on the IBM Quantum Experience.

\emph{Related work.} Several quantum programming languages, simulators, and compilers have been proposed and implemented in various languages. Yet, only a few of them are freely available: Quipper~\cite{green2013quipper}, a quantum program compiler implemented in Haskell, the ScaffCC compiler~\cite{javadiabhari2014scaffcc} based on the LLVM framework, and the LIQ$Ui\Ket{}$ simulator, which is implemented in F\#~\cite{wecker2014liqui} and only available as a binary. While all of these tools are important for the development of quantum computing in their own right, they have not (yet) been developed to a complete and unified software stack such as ProjectQ, containing the means to compile, simulate, emulate, and, ultimately, run quantum algorithms on actual hardware.

\emph{Outline.} After introducing the ProjectQ framework in Sec. \ref{sec:framework}, we motivate the methodology behind it in Sec.~\ref{sec:quantumprograms} using Shor's algorithm for factoring as an example. We then introduce the main features of our framework, including the high-level language, the compiler, and various back-ends in Sec.~\ref{sec:features}. Finally, we provide the road map for future extensions of the ProjectQ framework in Sec.~\ref{sec:roadmap}, which includes quantum chemistry and math libraries, support for further hardware and software back-ends, and additional compiler components.

\section{The ProjectQ Framework}
\label{sec:framework}
ProjectQ is an extensible open source software framework for quantum computing, providing clean interfaces for extending and improving its components. ProjectQ is built on four core principles: open \& free, simple learning curve, easily extensible, and high code quality.

\emph{Open \& free:} To encourage wide use, ProjectQ is being released as an open source software under the Apache 2 license. This is one of the most permissive license models and allows, for example, free commercial use.

\emph{Simple learning curve:} ProjectQ is implemented in Python (supporting both versions 2 and 3) because of its simple learning curve. Python is already widely used in the quantum community and easier to  learn than C\texttt{++} or functional programming languages. We  make use of high-performance code written in C\texttt{++} for some of the computational high performance kernels, but hide them behind a Python interface.

\emph{Easily extensible:} ProjectQ is easily extensible due to the modular implementation of both compiler and back-ends. This allows users to easily adapt the compiler to support new gates or entirely different gate sets as will be shown in a later section.

\emph{High code quality:} ProjectQ's code base follows high industry standards,  including mandatory code-reviews, continuous integration testing (currently 99\% line coverage using unit tests in addition to functional tests), and an extensive code documentation.

The initial release of ProjectQ, which is described in this paper, implements powerful core functionalities, and we will continue to add new features on a regular basis. We encourage contributions to the ProjectQ framework from anyone: Implementations of quantum algorithms, new compiler engines, and interfaces to actual quantum hardware back-ends are welcome. For more information about contributing see Ref. \cite{projectqwebsite}.

\section{Quantum Programs and Compilation}\label{sec:quantumprograms}
We motivate the methodology behind ProjectQ by presenting an implementation of Shor's factoring algorithm \cite{shor1994algorithms} in a high-level language. Then, we show how, using only local optimization and rewriting rules, the automatically generated low-level code can be as efficient as the manually optimized one in~\cite{beauregard2002circuit}, although all subroutines were implemented as stand-alone components.

\begin{figure}[H]
	\centering
	\includegraphics[width=\linewidth]{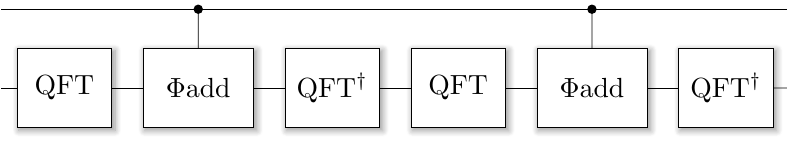}
	\caption{Circuit for carrying out two controlled Fourier transform additions in sequence: An optimizer can identify the QFT with its inverse, allowing to cancel those two operations. This identification is much harder or even impossible after decomposing operations and synthesizing rotation gates. Furthermore, since $\text{QFT}^\dagger \text{QFT}=\mathbbm 1$, the quantum Fourier transform does not need to be controlled, which can be achieved using our \texttt{Compute/Uncompute} meta-instructions (see Sec.~\ref{sec:metainstructions}).}
	\label{fig:twoadditions}
\end{figure}

\begin{figure*}[!t]
	\includegraphics[width=\linewidth]{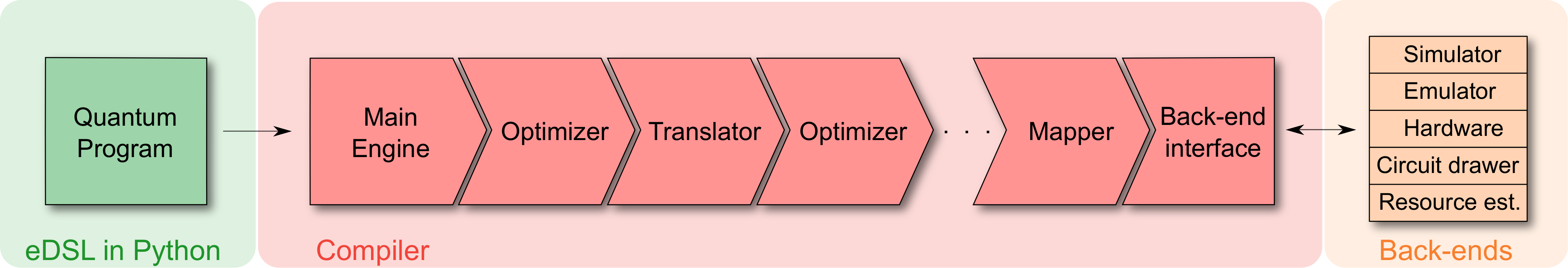}
	\caption{ProjectQ's full stack software framework. Users write their quantum programs in a high-level domain-specific language embedded in Python. The quantum program is then sent to the \texttt{MainEngine}, which is the front end of the modular compiler. The compiler consists of individual compiler engines which transform the code to the low-level instruction sets supported by the various back-ends, such as interfaces to quantum hardware, a high-performance quantum simulator and emulator, as well as a circuit drawer and a resource counter.}
	\label{fig:compilercomponents}
\end{figure*}

At the heart of Shor's algorithm for factoring~\cite{shor1994algorithms} lies modular exponentiation (of a classically-known constant by a quantum mechanical value) which is well-known to be implementable using constant-adders. Given a classical constant $c$, they transform a quantum register $\Ket x$ representing the integer $x$ as
\[
	\Ket x\mapsto\Ket{x+c}\;.
\]
Out of several implementations~\cite{takahashi2009quantum, haner2016factoring, draper2000addition}, we focus on Draper's addition in Fourier space~\cite{draper2000addition}, due to the large potential for optimization when executing several additions in sequence, which happens when using the construction in~\cite{beauregard2002circuit} to achieve modular addition. This constant-adder works as follows:
\begin{enumerate}[1.]
	\item Apply a quantum Fourier transform (QFT) to $\Ket x$.
	\item Apply phase-shift gates to the qubits of $\Ket x$, depending on the bit-representation of the constant $c$ to add (i.e., the addition is carried out in the phases).
	\item Apply an inverse QFT.
\end{enumerate}
Modular exponentiation can be implemented using repeated modular multiplication and shift, which themselves can be built using controlled modular adders. Thus, at a lower level, (double- or single-) controlled constant-adders are performed. When executing two such controlled additions in sequence, the resulting circuit can be optimized significantly. As shown in Fig.~\ref{fig:twoadditions}, the (final) inverse QFT of one addition and the initial QFT of the next one can be canceled. Furthermore, some of the (controlled) phase-shift gates inside $\Phi$add may be merged (depending on the two constants to add).

While the potential for these cancellations and optimizations is easy to see at this level of abstraction, carrying out such a cancellation is computationally very expensive once all gates have been translated to a low-level gate set, and may be impossible to do after translating into a discrete gate set (which introduces approximation errors). In our compilation framework, we thus define several intermediate gate sets. At every such intermediate level, inexpensive local optimization algorithms can be employed prior to further translation into the next lower-level representation.

The ProjectQ compiler is modular and allows new compilers to be built by combining existing and new components, as shown in Fig.~\ref{fig:compilercomponents}. This design allows to customize intermediate gate sets to improve optimization for specific algorithmic primitives. It also allows to adapt the compilation process to different quantum hardware architectures by replacing just some of the compiler engines (including hardware-specific mappers), which maximizes the re-use of individual compiler components.

\section{Features}\label{sec:features}

In this section we will introduce the main features of ProjectQ, starting with a minimal example:
\begin{lstlisting}[numbers=left,xleftmargin=21pt,framexleftmargin=17pt]
from projectq import MainEngine
from projectq.ops import H, Measure

eng = MainEngine()
qubit1 = eng.allocate_qubit()
H | qubit1
Measure | qubit1
print(int(qubit1))
\end{lstlisting}
This minimal code example allocates one qubit in state $\Ket 0$ and applies a Hadamard gate before measuring it in the computational basis and printing the outcome. While this is a valid quantum program implementing a random number generator, a more pythonic and better designed version is shown in code example \ref{code:first_program}.

\noindent Line 1 imports the \texttt{MainEngine} class, which is the front end of the quantum compiler as shown in Fig.~\ref{fig:compilercomponents}. Every quantum program needs to create one \texttt{MainEngine}, which contains all compiler components (engines) as well as the back-end (see line 4). In section~\ref{sec:compiler} we show how to select compiler engines and the back-ends. If a \texttt{MainEngine} is created without any arguments as done here, the default compiler engines are used with a simulator back-end.

Every quantum algorithm operates on qubits, which are obtained by calling the \texttt{allocate\_qubit} function of the \texttt{MainEngine}. To apply a Hadamard gate and then measure, we first need to import these gates in line 2 and apply them to the qubit in lines 6 and 7. The syntax for applying a quantum gate to a qubit mimics  an operator notation. For example,
\begin{equation}
	R_x(0.5) \Ket{\,\mathtt{qubit}\,} \;,
\end{equation}
might indicate a rotation around the $x$-axis applied to a qubit. In ProjectQ, this is coded as:
\begin{lstlisting}[numbers=none]
           Rx(0.5) | qubit
\end{lstlisting}
The symbol $|$ separates the gate operation with optional classical arguments (e.g., rotation angles) on the left  from its quantum arguments on the right (the qubits to which the gate is being applied).

Finally, on line 8, \texttt{qubit1} is converted to an $\texttt{int}$. This conversion operation returns the measurement result from line 7 which is then printed.

\begin{lstlisting}[float=*t,caption={Code example for implementing a quantum random number generator in our Python-embedded domain-specific language.},xleftmargin=21pt,framexleftmargin=17pt, label=code:first_program]
import projectq.setups.default               $\leftarrow$ # explicitly import default decompositions
from projectq import MainEngine              $\leftarrow$ # import the main compiler engine
from projectq.ops import H, Measure          $\leftarrow$ # import the required operations

def my_rng(eng):                             $\leftarrow$ # Python function definition
	qubit1 = eng.allocate_qubit()              $\leftarrow$ # allocate qubit1
	H | qubit1                                 $\leftarrow$ # apply a Hadamard gate to qubit1
	Measure | qubit1                           $\leftarrow$ # measure qubit1
	eng.flush()                                $\leftarrow$ # force-execute all gates
	return int(qubit1)                         $\leftarrow$ # access measurement result (conversion to int)

if __name__ == "__main__":                   $\leftarrow$ # only executes if this is main
	eng = MainEngine()                         $\leftarrow$ # create a main compiler engine
	print("Result: {}".format(my_rng(eng)))    $\leftarrow$ # call my_rng(eng) and print the result
\end{lstlisting}

\subsection{High level quantum language}
\label{sec:high_level_language}
\subsubsection{The basic (quantum) types}\label{sec:quantumtypes}
The fundamental types of our high level quantum language are logical qubits from which more complex types can built: quantum integers (quint), quantum fixed point numbers (qufixed) and quantum floats (qufloat). Similar to their classical counterparts, these types are just different interpretations of the contents of an underlying list of quantum bits, i.e., a quantum register or \texttt{qureg}.

To acquire an instance of an $n$-qubit quantum register, the function \texttt{allocate\_qureg(n)} of the \texttt{MainEngine} has to be invoked with the number of qubits as a parameter, as in the following code snippet:
\begin{lstlisting}[numbers=none]
...
eng = MainEngine()
...
qureg = eng.allocate_qureg(n)
\end{lstlisting}
The function \texttt{allocate\_qubit()} returns a quantum register of length one, which is a single qubit. Qubits can be allocated at any point in the program, which  keeps the user from having to specify a-priori the maximum number of qubits needed in any part of the code.

ProjectQ's compiler takes care of automatic deallocation of qubits by exploiting Python's garbage collection, thus allowing qubit re-use. While not necessary, the user can still force deallocation by invoking Python's \texttt{del} statement. In addition to removing the burden of resource management from the user, letting the compiler handle the life-time of qubits allows automatic parallelization for back-ends featuring more qubits than the minimal circuit width. The simulator can be used as a debugging tool to validate uncompute sections since it throws an exception whenever a qubit in superposition is being deallocated. Thus, qubits have to be either measured or uncomputed prior to deallocation.

Certain subroutines such as the multi-controlled NOT construction by Barenco et al.~\cite{barenco1995} or the constant-addition circuit by H\"aner et al.~\cite{haner2016factoring} do not require clean ancilla qubits in a defined computational basis state (such as $\ket{0}$) but work with borrowed qubits in an unknown arbitrary quantum state. They guarantee that after completion of the circuit, these so-called dirty ancilla qubits have returned to their starting state. Our compiler can thus optimize the allocation of such ancilla qubits by simply providing a qubit which is currently unused, independent of its state:
\begin{lstlisting}[numbers=none]
qubit = eng.allocate_qubit(dirty=True)
\end{lstlisting}

\subsubsection{Quantum gates and functions}
Operations on quantum data types can be implemented either as normal Python functions or as ProjectQ gates. A Python function implementing such an operation applies other gates or functions, as shown in this example of an addition by a constant:
\begin{lstlisting}[label={lst:addnocompuncomp},numbers=none]
def add_constant(quint, c):
	QFT | quint

	# addition in the phases:
	phi_add(quint, c)

	get_inverse(QFT) | quint
\end{lstlisting}
which can then be called as
\begin{lstlisting}[numbers=none]
add_constant(my_quint, 11)
\end{lstlisting}
The second approach is to define a custom ProjectQ gate representing the entire operation and then registering a decomposition rule, i.e., providing one possible function which can be used to replace the newly defined gate:
\begin{lstlisting}[label={lst:addgatedef},numbers=none]
class AddConstant(BasicGate):
	def __init__(self, c):
		self.c = c # store constant to add

# provide one possible decomposition:
register_decomposition(AddConstant, add_constant_decomposition1)
\end{lstlisting}
This enables the usual syntax for quantum gates, namely:
\begin{lstlisting}[numbers=none]
AddConstant(11) | my_quint
\end{lstlisting}
where the constant to add (11 in this case) is stored within the \texttt{AddConstant} gate object.

While defining a custom gate involves more code, it is superior to a function-based implementation and results in cleaner user code. The gate-based approach allows optimizations at a higher level of abstraction since the function call executes the individual operations right away, thus removing the potential of optimizing at the highest level. In our example, defining the constant-adder as a gate allows merging two consecutive \texttt{AddConstant} gates by adding the respective constants. This is much harder in the function-based implementation, where the highest level of abstraction which the compiler receives is at the level of \texttt{QFT} and phase-shift gates. 

Another advantage of implementing a new quantum operation, such as \texttt{AddConstant}, as a new ProjectQ gate is that different specialized decomposition rules can be defined from which the compiler can then choose the best one for the target back-end by evaluating a user-defined cost function. Furthermore, if the back-end natively supports certain gate operations, such as a many-qubit M{\o}lmer-S{\o}rensen gate~\cite{sorensen1999quantum} on ion trap quantum computers, the decomposition step may be skipped altogether. The \texttt{AddConstant} gate is also natively supported in our quantum emulator, which allows faster execution by orders of magnitude compared to simulating the individual gates of its implementation~\cite{haenerSC2016}.

Most generally, the quantum input to any gate is a tuple of quantum types (qubit, qureg, quint,...), which allows the following intuitive syntax for a \texttt{Multiply} instruction:
\begin{lstlisting}[numbers=none]
Multiply | (quint_a, quint_b, res)
\end{lstlisting}

\subsubsection{Meta-instructions}\label{sec:metainstructions}
At an even higher level of abstraction, complex gate operations can be modified using so-called \textit{meta-instructions}. These facilitate optimization processes in the compiler while allowing the user to write more concise code.

\begin{figure}
	\includegraphics[width=.7\linewidth]{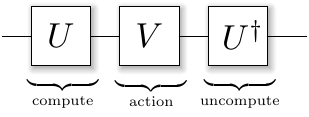}
	\caption{A compute/action/uncompute section. Since $U^\dagger U=\mathbbm 1$, the $U$ and $U^\dagger$ gate can be executed unconditionally when control qubits are added to an operation of this form. These sections can be annotated using our \texttt{Compute}, \texttt{Uncompute}, and \texttt{CustomUncompute} meta-instructions, see Sec.~\ref{sec:metainstructions}.}
	\label{fig:compuncomp}
\end{figure}

All of these meta-instructions are implemented using the Python context-handler syntax, i.e.,
\begin{lstlisting}[numbers=none]
with MetaInstructionObject:
	...
	...
\end{lstlisting}
where \texttt{MetaInstructionObject} is one of the following:
\begin{itemize}
	\item \texttt{Control(eng, control\_qubits)}: Condition an entire code block on one or several qubits being in state $\Ket 1$.
	\item \texttt{Compute(eng)/CustomUncompute(eng)}: Annotate compute / uncompute sections as depicted in Fig.~\ref{fig:compuncomp}. This allows to optimize the conditional execution of such sections as discussed in Ref.~\cite{haener2016}. For an automatic uncompute, simply run the \texttt{Uncompute(eng)} function.
	\item \texttt{Dagger(eng)}: Invert an entire unitary code block (hence the name: $U^\dagger U=\mathbbm 1$).
	\newpage
	\item \texttt{Loop(eng, num\_iterations)}: Run a code block \texttt{num\_iterations} times. Back-ends natively supporting loop instructions will receive the loop body only once, whereas the loop is unrolled otherwise. Some hardware platforms exploit this by re-using the generated waveforms.
\end{itemize}

For an example employing some of these meta-instructions to arrive at the performance-level of hand-optimized code, see our implementation of Shor's algorithm for factoring in the Appendix.

Any quantum program implemented in our eDSL concludes with the statement
\begin{lstlisting}[numbers=none]
eng.flush()
\end{lstlisting}
which makes sure that the entire circuit has passed through all compiler engines and is received by the back-end.

\subsection{Compiler and compiler engines}
\label{sec:compiler}

Our compiler is not a monolithic block, but has a modular design, allowing fast adaptation to new hardware while optimally re-using existing components. The compiler \texttt{MainEngine} serves as the front-end of the quantum compiler
and consists of a chain of compiler engines, each carrying out one specific task. This chain can be customized by the user. The most trivial compiler consists of one translation engine called \texttt{AutoReplacer}, which decomposes a quantum circuit into the native gate set of the back-end using the registered decomposition rules (see section \ref{sec:high_level_language}). Such a compiler can be instantiated as follows:
\begin{lstlisting}[numbers=none]
eng = MainEngine(engine_list=[AutoReplacer()])
\end{lstlisting}

Yet, as discussed in the previous section, optimizing at different levels of abstraction allows for more efficient code. Introducing multiple intermediate levels can be achieved using several translators, optimizers, and instruction filter engines, which define the set of supported gates at each level:

\begin{lstlisting}[numbers=none]
eng = MainEngine(engine_list=[AutoReplacer(),InstructionFilter(intermediate_gate_set),LocalOptimizer(),AutoReplacer(),LocalOptimizer()])
\end{lstlisting}

As quantum operations propagate through this compilation chain, the \texttt{AutoReplacer} engines decompose all instructions into the gate set defined by the next engine, which is, e.g., an \texttt{InstructionFilter}. The last engine is always the back-end, which defines the gate set at the lowest level.
For example, an  intermediate gate set supporting high-level QFT gates allows the optimization mentioned in Sec.~\ref{sec:quantumprograms} easily: consecutive (controlled) additions in Fourier space can be executed more efficiently by canceling a QFT with its inverse.

Once suitable compiler engines and decomposition rules have been determined for a specific back-end, those can be saved as a setup for future use.

Unlike other quantum compiler designs, our compiler does not store the entire quantum circuit which, given the vast number of logical gates required for some quantum algorithms, would not be feasible due to memory requirements. Instead, every compiler engine can define how many gates it stores. While, for example, an \texttt{AutoReplacer} works on only one gate at a time, a \texttt{LocalOptimizer} saves a short sequence of gates before trying to optimize them. This locality of compilation allows to parallelize the compilation process in a straight-forward manner.

\subsection{Back-ends}
Our compiler supports a wide range of back-ends to
\begin{itemize}
	\setlength{\itemsep}{5pt}
	\setlength{\parskip}{0pt}
	\setlength{\parsep}{0pt}
	\item run circuits on quantum hardware,
	\item simulate the individual gates of a quantum circuit
	\item emulate the action of a quantum circuit by employing high-level shortcuts
	\item estimate the required resources
	\item draw a quantum circuit
\end{itemize}
The default simulation back-end  can be changed by specifying the \texttt{backend}-parameter. To target the IBM Quantum Experience back-end instead, one simply writes
\begin{lstlisting}[numbers=none]
eng = MainEngine(backend=IBMBackend())
\end{lstlisting}

\subsubsection{Hardware back-end: IBM quantum experience}
\begin{figure*}[t]
\centering
\includegraphics[width=\linewidth]{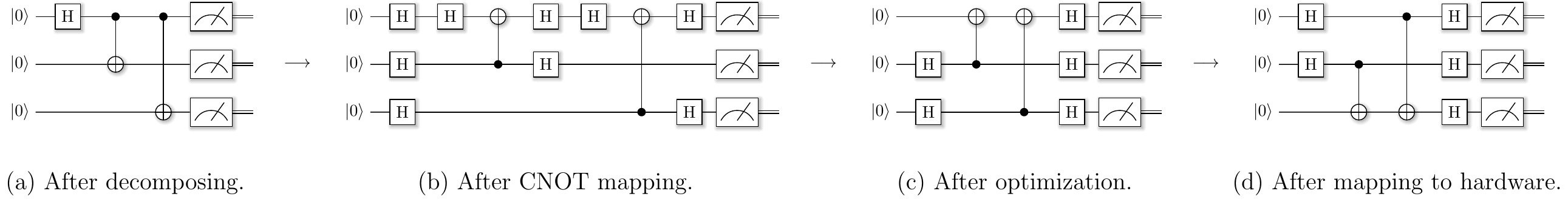}
\caption{Individual stages of compiling an entangling operation for the IBM back-end. The high-level Entangle-gate is decomposed into its definition (Hadamard gate on the first qubit, followed by a sequence of controlled NOT gates on all other qubits). Then, the CNOT gates are remapped to satisfy the logical constraint that controlled NOT gates are allowed to act on one qubit only, followed by optimizing and mapping the circuit to the actual hardware.}
\label{fig:ibmcompilation}
\end{figure*}

\begin{figure}[ht]
	\resizebox{\linewidth}{!}{\input{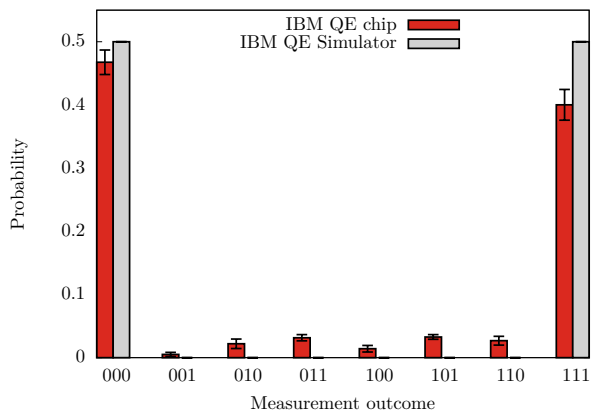}}
	\caption{Measurement outcomes with their respective probabilities when running an entangle operation on three qubits of the IBM Quantum Experience chip and simulator. The entangle operation corresponds to applying a Hadamard gate to the first qubit, followed by CNOT gates on all other qubits conditioned on the first qubit. The perfect outcome would be $50\%$ all-zeros and $50\%$ all-ones, as it is the case for the (noise-less) simulation.}
	\label{fig:ibmentanglehist}
\end{figure}

ProjectQ comes with a hardware backend for the IBM Quantum Experience device~\cite{ibmqewebsite}. As an example program, consider entangling $3$ qubits by first allocating $3$ qubits in a quantum register, then applying the \texttt{Entangle}-operation to them, and finally measuring the quantum register:
\begin{lstlisting}[numbers=none]
qureg = eng.allocate_qureg(3)
Entangle | qureg
Measure | qureg
eng.flush()
\end{lstlisting}
The compiler replaces the \texttt{Entangle} gate by a Hadamard gate on the first qubit, followed by CNOT gates on all others conditioned on the first qubit (see Fig.~\ref{fig:ibmcompilation}~a)). The compiler then automatically flips CNOT gates where necessary (using $4$ Hadamard gates), to make the code compatible with the connectivity of the IBM quantum chip (see Fig.~\ref{fig:ibmcompilation}~b)). In Fig.~\ref{fig:ibmcompilation}~c), the circuit is optimized before the final mapping to physical qubits is performed in Fig.~\ref{fig:ibmcompilation}~d). Running this circuit yields the outcomes with their respective probabilities in Fig.~\ref{fig:ibmentanglehist}. 

\subsubsection{Simulation of quantum circuits}

\begin{figure}[!h]
	\resizebox{\linewidth}{!}{
		\input{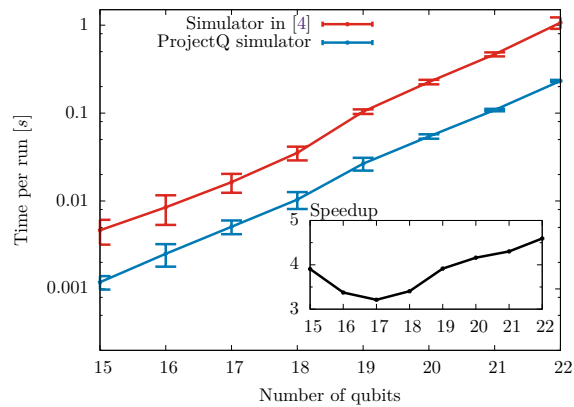}
	}
	\caption{Runtime comparison of the simulator from~\cite{haenerSC2016} to the simulator in ProjectQ. The timed circuit consists a Hadamard-transform, a chain of controlled z-rotations, and a final Hadamard transform. The lazy evaluation of gates in combination with intrinsics instructions allows the ProjectQ simulator to execute this circuit between 3 and 5 times faster. Both simulators were run on both cores of an Intel\textregistered{} Core\texttrademark{} i7-5600U CPU.}
	\label{fig:simcomparison}
\end{figure}

Simulating quantum programs at the level of individual gates can be achieved using our high-performance quantum simulator. The concrete gate set to be used can be specified by the user. The current version of the simulator in ProjectQ supports \texttt{AVX} instructions and OpenMP threads. Our simulator is substantially faster than the one we recently presented in Ref.~\cite{haenerSC2016}, which already outperfomed all other existing quantum simulators, see Fig.~\ref{fig:simcomparison}.

The simulation of quantum circuits at the level of gates is especially useful to simulate the effects of noise. We are working on the implementation of stochastic noise models and a high-performance density matrix simulator.

\subsubsection{Emulation of quantum circuits}

For efficient testing of quantum algorithms at a high level of abstraction, our simulator provides quantum emulation features as well: By specifying the level of abstraction at which to emulate (using, e.g., an \texttt{InstructionFilter} and an \texttt{AutoReplacer}, see Sec.~\ref{sec:compiler}), these features can be enabled or disabled. As an example, consider our \texttt{AddConstant} gate from Sec.~\ref{sec:high_level_language}: With a small modification to the gate definition, the addition can be carried out directly, without having to decompose it into QFT and phase-shift gates (and then further into 1- and 2-qubit gates only). Gates which execute classical mathematical functions on a superposition of values can derive from \texttt{BasicMathGate} and then provide a Python function mimicking this behavior in the \texttt{\_\_init\_\_} function of the \texttt{AddConstant} gate, i.e.,
\begin{lstlisting}[numbers=none]
BasicMathGate.__init__(self,
	lambda x: return x + c)
\end{lstlisting}

Such shortcuts allow faster execution by orders of magnitude, especially if potential low-level implementations require many ancilla qubits to perform the computation. These extra qubits do not need to be simulated when using this kind of shortcut, which allows to factor the number 
\[
	4,028,033 = 2,003\cdot 2,011
\]
on a regular laptop in less than 3 minutes.

The emulation of quantum circuits is very useful for determining mesh-sizes, time steps/slices, and other high-level parameters for, e.g., quantum chemistry applications~\cite{babbush2016exponentially}, which require many mathematical functions such as $\arcsin(x)$, $\exp(x)$, $\frac 1{\sqrt x}$, etc., to be evaluated on a superposition of values. The ProjectQ emulation feature then allows to perform numerical studies using the same code that was used to obtain resource estimates; and this can be achieved by merely changing one line of code.

\subsubsection{Resource estimation}

The \texttt{ResourceCounter} back-end can be inserted at any point in the compiler chain. There, it keeps track of all gates it encounters and measures the maximal circuit width at that level as well. In order to get resource estimates for very large circuits, caching must be introduced at every intermediate layer / gate set. This feature will be added in the near future.

\subsubsection{Circuit drawing back-end}
\begin{figure}[tb]
	\includegraphics[width=\linewidth]{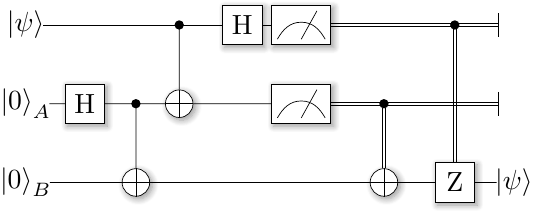}
	\caption{Quantum circuit for quantum teleportation. This circuit was automatically generated (modulo renaming of the qubits).}
	\label{fig:teleportcirc}
\end{figure}

Last but not least, all quantum programs can be drawn as circuit diagrams for publication in, e.g., an algorithmic paper. The \texttt{CircuitDrawer} back-end saves the circuit and, upon invocation of the \texttt{get\_latex()} member function, returns TikZ-\LaTeX code ready to be published. See Fig.~\ref{fig:teleportcirc} for an example output depicting quantum teleportation (including the initial creation of a Bell-pair). Just like the simulator or resource counter, this back-end can also be used as an intermediate compiler engine in order to draw the circuit at various levels of abstraction (which was done to create Fig.~\ref{fig:ibmcompilation}). This back-end will also be extended with further features in future versions of ProjectQ, giving the user yet more power over the mapping \& drawing process.
\vspace{10pt}

\section{Road map}\label{sec:roadmap}

In order to widen the scope of available tools, we will be extending ProjectQ on a regular basis with libraries, additional compiler engines, and more hardware back-ends.

\subsection{Libraries}
\noindent\emph{fermilib}. Solving problems involving strongly interacting fermions is one of the most promising applications for near-term quantum devices. With external collaborators we are implementing \texttt{fermilib} \cite{fermilib}, a library for designing quantum simulation algorithms to treat fermionic systems. \texttt{fermilib} will include integration with open source electronic structure packages to enable the computation of arbitrary molecular Hamiltonians. A well-engineered Python interface will allow for efficient manipulation of fermionic data structures with routines enabling normal ordering, fast orbital transformations, mapping qubit algebras, and more. In addition, it will provide tools to study fermionic systems beyond chemistry, including models for superconductivity such as the Hubbard model.

\vspace{5pt}
\noindent\emph{mathlib}. While our emulator can use classical shortcuts to mimic the application of general mathematical functions, manually-tuned high-performance implementations of those functions at the level of, e.g., Toffoli gates are still required in order to get resource estimates or, at a later point in time, run those algorithms on real quantum hardware. We will thus further extend the (small) existing math library in ProjectQ with additional high-level quantum types (as discussed in Sec.~\ref{sec:high_level_language}) and corresponding operations.

\subsection{Back-ends}

We are working on adding support for further hardware back-ends. Among others, we aim to have an interface to J. Home's ion trap devices \cite{de2016parallel} in the near future. More hardware back-ends will follow soon. Hardware groups interested in interfacing to ProjectQ are highly encouraged to contact us~\cite{projectqwebsite}.

Also, we will keep extending and improving our classical back-ends. We are currently working on including our distributed massively parallel quantum simulator \cite{haener2017}, which allows to simulate up to 45 qubits on one of the world's largest supercomputers.

\subsection{Compiler extensions}

New compiler engines will be added, allowing to deal with more advanced layouting, which is required to employ quantum error-correction schemes. This is crucial not only to run large circuits on future hardware, but also to gain information about resource usage before large-scale quantum computers are available.

\acknowledgments

Special thanks go to 
the researchers from IBM,
Lev S. Bishop,
Fran Cabrera,
Jorge Carballo,
Jerry M. Chow,
Andrew W. Cross,
Ismael Faro,
Stefan Filipp,
Jay M. Gambetta,
Paco Martin,
Nikolaj Moll,
Mark Ritter, and
John Smolin
for their help with interfacing to the IBM Quantum Experience chip.

We thank 
Jonathan Home,
Matteo Marinelli, and
Vlad Negnevitsky
for working with us on an interface to their ion trap quantum computer.

Furthermore, we want to thank the following people for enlightening discussions:
Jana Darulov\'a,
Michele Dolfi,
Dominik Gresch,
Andreas Hehn,
Mario K\"onz,
Natalie Pearson,
Donjan Rodic,
Slava Savenko,
Andreas Wallraff, and
Camillo Zapata Ocampo
from ETH Zurich;
Matthew Neeley,
Daniel Sank, and
Hartmut Neven
from Google Quantum AI;
Alan Geller,
Martin Roetteler,
Krysta Svore,
Dave Wecker, and
Nathan Wiebe
from Microsoft Research;
and
Anne Matsuura and
Mikhail Smelyanskiy
from Intel.

We would also like to thank
Ryan Babbush,
Jarrod McClean, and
Ian D. Kivlichan
for collaborating with us on \texttt{fermilib}.

We acknowledge support by the Swiss National Science Foundation and the Swiss National Competence Center for Research, QSIT.

\bibliographystyle{unsrtnat}
\bibliography{references}

\appendix
\newpage
\section{Examples}
In this section, we will discuss complete examples, which are also included with the ProjectQ sources.

\subsection{Quantum Teleportation}

\noindent Quantum teleportation can be implemented as follows:
\begin{lstlisting}[numbers=none]
# allocate 2 qubits and turn them into a Bell-pair (entangle them)
b1 = eng.allocate_qubit()
b2 = eng.allocate_qubit()
H | b1
CNOT | (b1, b2)

# Alice creates a nice state to send
psi = eng.allocate_qubit()
create_state(psi)

# entangle it with Alice's b1
CNOT | (psi, b1)

# measure two values (once in Hadamard basis) and send the bits to Bob
H | psi
Measure | (psi, b1)
print("Message: {}".format([int(psi), int(b1)])

# Bob may have to apply up to two operation depending on the message sent by Alice:
with Control(eng, b1):
	X | b2
with Control(eng, psi):
	Z | b2
# done.
\end{lstlisting}
When using the \texttt{CircuitDrawer} back-end, i.e.,
\begin{lstlisting}[numbers=none]
eng = MainEngine(CircuitDrawer())
\end{lstlisting}
the quantum circuit depicted in Fig.~\ref{fig:teleportcirc} is generated.

\subsection{Grover search}

Grover's search algorithm~\cite{grover1996fast} achieves a quadratic speedup for finding an element $e$, given a function
\[
	f(x)=\left\{\begin{matrix}
	1,\; x = e\\
	0,\; x \neq e
	\end{matrix}\right.
\]
It requires two quantum oracles to be implemented; one is $U_f$, which marks the element $e$ by adding a phase of $-1$ to the quantum state:
\begin{align*}
	U_f\Ket x &= \Ket x,\;x\neq e\\
	U_f\Ket e &= -\Ket e
\end{align*}
and the other oracle is a reflection across the uniform superposition. These two operators are then applied iteratively $\frac{\pi\sqrt N}4$ times, where $N$ is the number of potential inputs to the function.

An example implementation of this algorithm with $e=1010...101_2$ (the binary representation of the solution is alternating) is available in the examples folder of the ProjectQ framework. The loop performing the two reflections looks as follows:
\begin{lstlisting}[numbers=none]
# run num_it iterations
with Loop(eng, num_it):
	# adds a (-1)-phase to the solution
	oracle(eng, x, oracle_out)
		
	# reflection across uniform superposition:
	# map uniform superposition to all-ones
	with Compute(eng):
		All(H) | x
		All(X) | x
	# phase-flip for all-ones:
	with Control(eng, x[0:-1]):
		Z | x[-1]
	# undo mapping
	Uncompute(eng)
\end{lstlisting}
Where the \texttt{Loop} meta-instruction does not unroll the loop if the underlying hardware or further compiler engines support the execution or optimization of loops.

\newpage
\subsection{Shor's algorithm for factoring}
Our eDSL allows the user to implement more complex functions nicely while keeping the efficiency of the resulting code at the level of a hand-optimized implementation. As an example, consider the modular adder proposed by Beauregard~\cite{beauregard2002circuit}, which can be implemented using our powerful meta-instructions:
\begin{lstlisting}[numbers=none]
def add_constant_modN(eng, c, N, quint):
	assert(c < N and c >= 0)

	AddConstant(c) | quint

	with Compute(eng):
		SubConstant(N) | quint
		ancilla = eng.allocate_qubit()
		CNOT | (quint[-1], ancilla)
		with Control(eng, ancilla):
			AddConstant(N) | quint

	SubConstant(c) | quint

	with CustomUncompute(eng):
		X | quint[-1]
		CNOT | (quint[-1], ancilla)
		X | quint[-1]
		del ancilla

	AddConstant(c) | quint
\end{lstlisting}

A complete implementation of Shor's algorithm is also available in the examples folder of ProjectQ. Using the quantum math library, an implementation of this algorithm takes only a few lines. The iterative modular multiplication and shift, which is used to implement the modular exponentiation of $a$ by a $2n$ bit quantum number $x$ in a uniform superposition looks as follows:
\newpage
\begin{lstlisting}[numbers=none]
# do each of the bits of x separately, employing the semi-classical inverse QFT
ctrl_qubit = eng.allocate_qubit()

# loop over all 2n bits
for k in range(2 * n):
	# shift the a we multiply by
	current_a = pow(a, 1 << (2 * n - 1 - k), N)
	
	# x is in uniform superposition:
	H | ctrl_qubit
	
	# apply controlled modular multiplication
	with Control(eng, ctrl_qubit):
		MultiplyByConstantModN(current_a, N) | x

	# perform inverse QFT -> Rotations conditioned on previous outcomes
	for i in range(k):
		if measurements[i]:
			R(-math.pi/(1 << (k - i))) | ctrl_qubit
	
	# final Hadamard of the inverse QFT
	H | ctrl_qubit

	# and measure
	Measure | ctrl_qubit
	eng.flush()
	
	# store the measurement result
	measurements[k] = int(ctrl_qubit)
	
	# and reset the qubit for the next iteration
	if measurements[k]:
		X | ctrl_qubit
\end{lstlisting}

\end{document}